\documentclass[letterpaper, 10 pt, conference]{ieeeconf}  

\usepackage{amsmath}
\usepackage{graphicx}
\usepackage{cite}
\usepackage{balance}
\usepackage{amssymb,amsfonts}

\usepackage{amssymb,amsfonts}

\usepackage{lipsum}
\usepackage{mathtools}
\usepackage{cuted}

\usepackage{balance}

\newtheorem{theorem}{\it Theorem}
\newtheorem{lemma}{\it Lemma}
\newtheorem{definition}{\it Definition}

\newtheorem{corollary}{\it Corollary}

\IEEEoverridecommandlockouts                              

\title{\LARGE \bf
	Fundamental Stealthiness-Distortion Tradeoffs in Dynamical Systems under Injection Attacks: A Power Spectral Analysis
}

\author{Song Fang$^{1}$ and Quanyan Zhu$^{1}$
	\thanks{This research is partially supported by award 2015-ST-061-CIRC01 from U.S. DHS, awards ECCS-1847056, CNS-1544782, CNS-2027884, and SES-1541164 from NSF, and grant W911NF-19-1-0041 from ARO.}
	\thanks{$^{1}$ Song Fang and Quanyan Zhu are with the Department of Electrical and Computer Engineering, New York University, New York, USA
		{\tt\small song.fang@nyu.edu; quanyan.zhu@nyu.edu}}%
}

\begin{document}

	\maketitle
	\thispagestyle{empty}
	\pagestyle{empty}

	\begin{abstract}
		
		In this paper, we analyze the fundamental stealthiness-distortion tradeoffs of linear Gaussian dynamical systems  under data injection attacks using a power spectral analysis, whereas the Kullback--Leibler (KL) divergence is employed as the stealthiness measure. Particularly, we obtain explicit formulas in terms of power spectra that characterize analytically the stealthiness-distortion tradeoffs as well as the properties of the worst-case attacks. Furthermore, it is seen in general that the attacker only needs to know the input-output behaviors of the systems in order to carry out the worst-case attacks.

\end{abstract}


\section{Introduction}

The Kullback--Leibler (KL) divergence was proposed in \cite{kullback1951information} (see also \cite{kullback1997information}), and has been employed in various research areas, including, e.g., information theory \cite{Cov:06}, signal processing \cite{Kay20}, statistics \cite{pardo2006statistical}, control and estimation theory \cite{lindquist2015linear}, system identification \cite{stoorvogel1996system}, and machine learning \cite{goodfellow2016deep}. Particularly, in detection theory \cite{poor2013introduction}, KL divergence provides the optimal exponent in probability of error for binary hypotheses testing problems \cite{Cov:06}, for instance, regarding whether an attack signal is present or not in security problems; accordingly, the KL divergence has been employed as a measure of stealthiness for attacks (see detailed discussions in, e.g., \cite{bai2017kalman, bai2017data}).

In the context of dynamical control security (see, e.g., \cite{poovendran2012special, johansson2014guest, sandberg2015cyberphysical, cheng2017guest, giraldo2018survey, weerakkody2019resilient, dibaji2019systems, chong2019tutorial} and the references therein), particularly in dynamical control systems under injection attacks, fundamental stealthiness-distortion tradeoffs (with the KL divergence being the stealthiness measure) have been investigated for feedback control systems (see, e.g., \cite{zhang2017stealthy, bai2017data}) as well as state estimation systems (see, e.g., \cite{bai2017kalman, kung2016performance, guo2018worst}).
Generally speaking, the problem considered is: Given a constraint (upper bound) on the level of stealthiness, what is the maximum degree of distortion (for control or for estimation) that can be caused by the attacker? This is dual to the following question: Given a least requirement (lower bound) on the degree of distortion, what is the minimum level of stealthiness that can be achieved by the attacker? Answers to these questions can not only capture the fundamental tradeoffs between stealthiness and distortion, but also characterize what the worst-case attacks are.


In this paper, unlike the aforementioned works in \cite{bai2017kalman, kung2016performance, zhang2017stealthy, bai2017data, guo2018worst}, we adopt an alternative approach to this stealthiness-distortion tradeoff problem using power spectral analysis. The scenario we consider is linear Gaussian 
dynamical systems.
By using the power spectral approach, we obtain explicit formulas that characterize analytically the stealthiness-distortion tradeoffs as well as the properties of the worst-case attacks. 
It turns out that the worst-case attacks are stationary colored Gaussian attacks with power spectra that are shaped specifically according to the transfer functions of the systems and the power spectra of the system outputs, the knowledge of which is all that the attacker needs to have access to in order to carry out the worst-case attacks.


The remainder of the paper is organized as follows. Section~II provides the technical preliminaries. Section~III presents the main results. Concluding remarks are given in Section~IV.

Note that an arXiv version of this paper with additional results and discussions can be found in \cite{KLarxiv}.

\section{Preliminaries}

Throughout the paper, we consider zero-mean real-valued continuous random variables and random vectors, as well as discrete-time stochastic processes. We represent random variables and random vectors using boldface letters, e.g., $\mathbf{x}$, while the probability density function of $\mathbf{x}$ is denoted as $p_\mathbf{x}$. In addition, $\mathbf{x}_{0,\ldots,k}$ will be employed to denote the sequence $\mathbf{x}_{0}, \ldots, \mathbf{x}_{k}$ or the random vector $\left[ \mathbf{x}_0^{\mathrm{T}},\ldots,\mathbf{x}_{k}^{\mathrm{T}} \right]^{\mathrm{T}}$, depending on the context. Note in particular that, for simplicity and with abuse of notations, we utilize $\mathbf{x} \in \mathbb{R}$ and $\mathbf{x} \in \mathbb{R}^m$ to indicate that $\mathbf{x}$ is a real-valued random variable and that $\mathbf{x}$ is a real-valued $m$-dimensional random vector, respectively.

A stochastic process $\left\{ \mathbf{x}_{k}\right\}, \mathbf{x}_k \in  \mathbb{R}$ is said to be  stationary if $ R_{\mathbf{x}}\left( i,k\right)=\mathbb{E}\left[  \mathbf{x}_i \mathbf{x}_{i+k} \right]$ depends only on $k$, and can thus be denoted as  $R_{\mathbf{x}}\left( k\right)$ for simplicity. The power spectrum of a stationary  process $\left\{ \mathbf{x}_{k} \right\}, \mathbf{x}_{k} \in \mathbb{R}$ is defined as
\begin{flalign}
S_{\mathbf{x}}\left( \omega\right)
=\sum_{k=-\infty}^{\infty} R_{\mathbf{x}}\left( k\right) \mathrm{e}^{-\mathrm{j}\omega k}. \nonumber
\end{flalign}
Moreover, the variance of $\left\{ \mathbf{x}_{k}\right\}$ is given by
\begin{flalign}
\sigma_{\mathbf{x}}^2
= \mathbb{E}\left[ \mathbf{x}_k^2 \right]
= \frac{1}{2\pi}\int_{-\pi}^{\pi} S_{\mathbf{x}}\left(\omega \right) \mathrm{d}  \omega. \nonumber
\end{flalign}

The KL divergence (see, e.g., \cite{kullback1951information}) is defined as follows.

\begin{definition}
	Consider random vectors $\mathbf{x} \in \mathbb{R}^m$ and $\mathbf{y} \in \mathbb{R}^m$ with probability densities $p_\mathbf{x} \left( \mathbf{u} \right)$ and $p_\mathbf{y} \left( \mathbf{u} \right)$, respectively.
	The KL divergence from distribution $p_\mathbf{x}$
	to distribution $p_\mathbf{y}$ is defined as
	\begin{flalign}
	\mathrm{KL} \left(p_{\mathbf{y}} \| p_{\mathbf{x}} \right)
	= \int p_{\mathbf{y}} \left( \mathbf{u} \right) \ln \frac{p_{\mathbf{y}} \left( \mathbf{u} \right)}{ p_{ \mathbf{x} } \left( \mathbf{u} \right)} \mathrm{d} \mathbf{u}. \nonumber
	\end{flalign}
\end{definition}

\vspace{3mm}

The next lemma (see, e.g., \cite{Kay20}) provides an explicit expression of KL divergence in terms of covariance matrices for Gaussian random vectors; note that herein and in the sequel, all random variables and random vectors are assumed to be zero-mean.

\begin{lemma} \label{Gaussian}
	Consider Gaussian random vectors $\mathbf{x} \in \mathbb{R}^m$ and $\mathbf{y} \in \mathbb{R}^m$ with covariance matrices $\Sigma_\mathbf{x}$ and $\Sigma_\mathbf{y}$, respectively.
	The KL divergence from distribution $p_\mathbf{x}$
	to distribution $p_\mathbf{y}$ is given by
	\begin{flalign}
	\mathrm{KL} \left( p_{\mathbf{y}} \| p_{\mathbf{x}} \right) = \frac{1}{2} \left[ \mathrm{tr} \left( \Sigma_{\mathbf{y}} \Sigma_{\mathbf{x}}^{-1} \right) - \ln \det \left( \Sigma_{\mathbf{y}} \Sigma_{\mathbf{x}}^{-1} \right) - m \right].
	\nonumber
	\end{flalign}
\end{lemma}

\vspace{3mm}

It is clear that in the scalar case (when $m=1$), Lemma~\ref{Gaussian} reduces to the following formula for Gaussian random variables:
\begin{flalign}
\mathrm{KL} \left( p_{\mathbf{y} } \| p_{\mathbf{x} } \right) = \frac{1}{2} \left[  \frac{ \sigma_{\mathbf{y} }^2}{\sigma_{\mathbf{x} }^2} 
-  \ln  \left( \frac{ \sigma_{\mathbf{y} }^2 }{\sigma_{\mathbf{x} }^2} \right) - 1 \right].
\nonumber
\end{flalign}

The KL divergence rate (see, e.g., \cite{lindquist2015linear}) is defined as follows.

\begin{definition}
	Consider stochastic processes $\left\{ \mathbf{x}_k \right\}, \mathbf{x}_k \in \mathbb{R}$ and $\left\{ \mathbf{y}_k \right\}, \mathbf{y}_k \in \mathbb{R}$ with densities $p_\mathbf{\left\{ \mathbf{x}_k \right\}}$ and $p_\mathbf{\left\{ \mathbf{y}_k \right\}}$, respectively; note that  $p_\mathbf{\left\{ \mathbf{x}_k \right\}}$ and $p_\mathbf{\left\{ \mathbf{y}_k \right\}}$ will be denoted by $p_\mathbf{x}$ and $p_\mathbf{y}$ for simplicity in the sequel.
	Then, the KL divergence rate from distribution $p_\mathbf{x}$
	to distribution $p_\mathbf{y}$ is defined as
	\begin{flalign}
	\mathrm{KL}_{\infty} \left( p_{\mathbf{y}} \| p_{\mathbf{x}} \right)
	= \limsup_{k \to \infty} \frac{\mathrm{KL} \left( p_{\mathbf{y}_{0, \ldots, k}} \| p_{\mathbf{x}_{0, \ldots, k}} \right)}{k+1}. \nonumber
	\end{flalign}
\end{definition}

\vspace{3mm}

The next lemma (see, e.g., \cite{lindquist2015linear}) provides an explicit expression of KL divergence rate in terms of power spectra for stationary Gaussian processes.

\begin{lemma} 
	Consider stationary Gaussian processes $\left\{ \mathbf{x}_k \right\}, \mathbf{x}_k \in \mathbb{R}$ and $\left\{ \mathbf{y}_k \right\}, \mathbf{y}_k \in \mathbb{R}$ with densities $p_\mathbf{x}$ and $p_\mathbf{y}$ as well as power spectra $S_\mathbf{x} \left( \omega \right)$ and $S_\mathbf{y} \left( \omega \right)$, respectively. Suppose that 
	$S_{\mathbf{y}} \left(\omega \right) / S_{\mathbf{x}} \left( \omega \right)$ is bounded (see \cite{lindquist2015linear} for details). Then, the KL divergence rate from distribution $p_\mathbf{x}$
	to distribution $p_\mathbf{y}$ is given by
	\begin{flalign} \label{ISD}
	&\mathrm{KL}_{\infty} \left( p_{\mathbf{y}} \| p_{\mathbf{x}} \right) \nonumber \\
	&~~~~ =  \frac{1}{2 \pi} \int_{0}^{2 \pi} \frac{1}{2} \left\{ \frac{S_{\mathbf{y}} \left(\omega \right)}{S_{\mathbf{x}} \left( \omega \right)} - \ln \left[ \frac{S_{\mathbf{y}} \left(\omega \right)}{S_{\mathbf{x}} \left( \omega \right)} \right] - 1 \right\} \mathrm{d} \omega.
	\end{flalign}
\end{lemma}

\vspace{3mm}

In fact, the right-hand side of \eqref{ISD} is also known as (half of) the Itakura--Saito distance, which is defined as
\begin{flalign} 
&\mathrm{D}_{\mathrm{IS}} \left( S_{\mathbf{y}} \left(\omega \right) , S_{\mathbf{x}} \left(\omega \right) \right) \nonumber \\
&~~~~ =  \frac{1}{2 \pi} \int_{0}^{2 \pi} \left\{ \frac{S_{\mathbf{y}} \left(\omega \right)}{S_{\mathbf{x}} \left( \omega \right)} - \ln \left[ \frac{S_{\mathbf{y}} \left(\omega \right)}{S_{\mathbf{x}} \left( \omega \right)} \right] - 1 \right\} \mathrm{d} \omega. \nonumber
\end{flalign} 
See also, e.g., \cite{georgiou2008metrics, ferrante2012time}, for further details of this metric.

\vspace{3mm}

\section{Stealthiness-Distortion Tradeoffs and Worst-Case Attacks}

In this section, we analyze the fundamental
stealthiness-distortion tradeoffs of linear Gaussian dynamical systems under data injection attacks, whereas the KL
divergence is employed as the stealthiness measure. Consider the scenario where attacker can modify the system input, and consequently, the system state and system output will then all be changed. From the attacker's point of view, the desired outcome is that the change in system state (as measured by state distortion) is large while the change in system output (as measured by output stealthiness) is relatively small, so as to make the possibility of being detected low. Meanwhile fundamental tradeoffs in general exist between state distortion and output stealthiness, since the system's state and output are correlated. In other words, increase in state distortion may inevitably lead to decrease in output stealthiness, i.e., increase in the possibility of being detected. How to capture such tradeoffs? And what is the worst-case attack that can cause the maximum distortion given a certain stealthiness level, or vice versa? The answers are provided subsequently in terms of power spectral analysis.


\begin{figure}
	\vspace{-3mm}
	\begin{center}
		\includegraphics [width=0.5\textwidth]{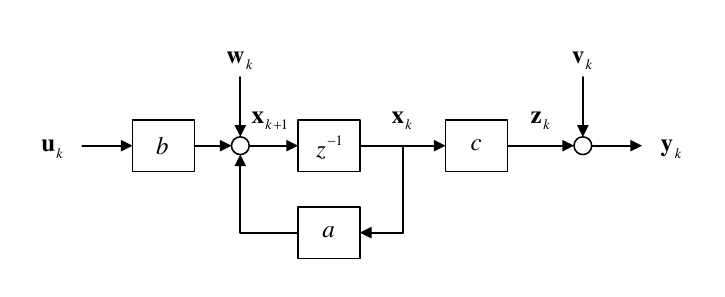}
		\vspace{-9mm}
		\caption{A Dynamical System.}
		\label{system1}
	\end{center}
	\vspace{-6mm}
\end{figure}


More specifically, consider the scalar dynamical system depicted in Fig.~\ref{system1}  with state-space model given by
\begin{flalign}
\left\{ \begin{array}{rcl}
\mathbf{x}_{k+1} & = & a \mathbf{x}_{k} + b \mathbf{u}_{k} + \mathbf{w}_k,\\
\mathbf{y}_{k} & = & c \mathbf{x}_{k} +\mathbf{v}_k,
\end{array} \right. \nonumber
\end{flalign}
where $\mathbf{x}_{k} \in \mathbb{R}$ is the system state, $\mathbf{u}_{k} \in \mathbb{R}$ is the system input, $\mathbf{y}_{k} \in \mathbb{R}$ is the system output, $\mathbf{w}_{k} \in \mathbb{R}$ is the process noise, and $\mathbf{v}_{k} \in \mathbb{R}$ is the measurement noise.
The system parameters are $ a \in \mathbb{R}$, $ {b} \in \mathbb{R}$, and $ {c} \in \mathbb{R}$; we further assume that $\left| a \right| < 1$ and $b,c \neq 0$, i.e., the system is stable, controllable, and observable. Accordingly, the transfer function of the system is given by
\begin{flalign}
P \left( z \right)
= \frac{bc}{z - a}. \nonumber
\end{flalign} 
(It is clear that $P \left( z \right)$ is minimum-phase.)
Suppose that $\left\{ \mathbf{w}_{k} \right\}$ and $\left\{ \mathbf{v}_{k} \right\}$ are stationary white Gaussian with variances  $\sigma_{\mathbf{w}}^2$ and $\sigma_{\mathbf{v}}^2$, respectively. Furthermore,  $\left\{ \mathbf{w}_{k} \right\}$, $\left\{ \mathbf{v}_{k} \right\}$, and $\mathbf{x}_{0}$ are assumed to be mutually independent. Assume also that $\left\{ \mathbf{u}_{k} \right\}$ is stationary with power spectrum $S_{\mathbf{u}} \left(\omega \right) $. As such, $\left\{ \mathbf{x}_{k} \right\}$ and $\left\{ \mathbf{y}_{k} \right\}$ are both stationary, and denote their power spectra by $S_{\mathbf{x}} \left(\omega \right) $ and $S_{\mathbf{y}} \left(\omega \right) $, respectively.

\begin{figure}
	\vspace{-3mm}
	\begin{center}
		\includegraphics [width=0.5\textwidth]{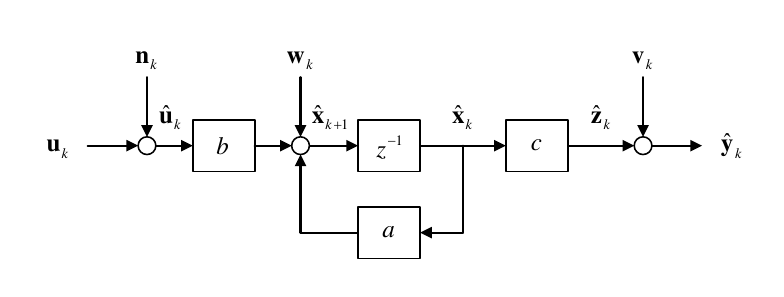}
		\vspace{-9mm}
		\caption{A Dynamical System under Injection Attack.}
		\label{system2}
	\end{center}
	\vspace{-6mm}
\end{figure}

Consider then the scenario that an attack signal $\left\{ \mathbf{n}_{k} \right\}, \mathbf{n}_{k} \in \mathbb{R}$, is to be added to the input of the system $\left\{ \mathbf{u}_{k} \right\}$ to deviate the system state $\left\{ \mathbf{x}_{k} \right\}$, while aiming to be stealthy in the system output; see the depiction in Fig.~\ref{system2}. Suppose that the attack signal $\left\{ \mathbf{n}_{k} \right\}$ is independent of $\left\{ \mathbf{u}_{k} \right\}$, $\left\{ \mathbf{w}_{k} \right\}$, $\left\{ \mathbf{v}_{k} \right\}$, and $\mathbf{x}_{0}$; consequently, $\left\{ \mathbf{n}_{k} \right\}$ is independent of $\left\{ \mathbf{x}_{k} \right\}$ and $\left\{ \mathbf{y}_{k} \right\}$ as well. 
The following questions then naturally arise: What is the fundamental tradeoff between the degree of distortion caused in the system state and the level of stealthiness resulted in the system output? More specifically, to achieve a certain degree of distortion in state, what is the minimum level of stealthiness that can be maintained by the attacker? And what is the worst-case attack in this sense?

The following theorem, as the first main result of this paper, answers the questions raised above.

\begin{theorem} \label{t1}
	Consider the dynamical system under injection attacks depicted in Fig.~\ref{system2}. Suppose that the attacker aims to design the attack signal $\left\{ \mathbf{n}_{k} \right\}$ to satisfy the following attack goal in terms of state distortion:  
	\begin{flalign}
	\mathbb{E} \left[ \left( \widehat{\mathbf{x}}_k - \mathbf{x}_{k} \right)^{2} \right] \geq D.
	\end{flalign} 
	Then, the minimum KL divergence rate between the original output and the attacked output is given by
	\begin{flalign} \label{leakage1}
	&\inf_{\mathbb{E} \left[ \left( \widehat{\mathbf{x}}_k - \mathbf{x}_{k} \right)^{2} \right] \geq D} \mathrm{KL}_{\infty} \left( p_{\widehat{\mathbf{y}}} \| p_{\mathbf{y}} \right) \nonumber \\
	&~~~~ = \frac{1}{2 \pi} \int_{0}^{2 \pi} \frac{1}{2} \left\{ \frac{S_{\widehat{\mathbf{n}}} \left(\omega \right)}{S_{\mathbf{y}} \left( \omega \right)} - \ln \left[ 1 + \frac{S_{\widehat{\mathbf{n}}} \left(\omega \right)}{S_{\mathbf{y}} \left( \omega \right)} \right] \right\} \mathrm{d} \omega,
	\end{flalign} 
	where
	\begin{flalign}
	S_{\widehat{\mathbf{n}}} \left(\omega \right) 
	= \frac{\zeta S_{\mathbf{y}}^2 \left( \omega \right)}{ 1 - \zeta S_{\mathbf{y}} \left( \omega \right)},
	\end{flalign}
	and $S_{\mathbf{y}} \left( \omega \right)$ is given by 
	\begin{flalign} \label{yspectrum}
	S_{\mathbf{y}} \left( \omega \right)
	&= \frac{b^2 c^2}{\left| \mathrm{e}^{\mathrm{j} \omega} - a \right|^2} S_{\mathbf{u}} \left( \omega \right)
	+ \frac{c^2}{\left| \mathrm{e}^{\mathrm{j} \omega} - a \right|^2} \sigma_{\mathbf{w}}^2 + \sigma_{\mathbf{v}}^2. 
	\end{flalign} 
	Herein, $\zeta$ is the unique constant that satisfies
	\begin{flalign}
	\frac{1}{2\pi} \int_{-\pi}^{\pi} \frac{\zeta S_{\mathbf{y}}^2 \left( \omega \right)}{ 1 - \zeta S_{\mathbf{y}} \left( \omega \right)} \mathrm{d}  \omega =c^2 D,
	\end{flalign}
	while
	\begin{flalign}
	0 < \zeta < \min_{\omega} \frac{1}{S_{\mathbf{y}} \left( \omega \right)}.
	\end{flalign}
	Moreover, the worst-case (in the sense of achieving this minimum KL divergence rate) attack
	$\left\{ \mathbf{n}_{k} \right\}$ is a stationary colored Gaussian process with power spectrum 	\begin{flalign}
	S_{\mathbf{n}} \left( \omega \right)
	= \frac{\left| \mathrm{e}^{\mathrm{j} \omega} - a \right|^2}{b^2 c^2} \frac{\zeta S_{\mathbf{y}}^2 \left( \omega \right)}{ 1 - \zeta S_{\mathbf{y}} \left( \omega \right)}.
	\end{flalign} 
\end{theorem}

\begin{figure}
	\vspace{-3mm}
	\begin{center}
		\includegraphics [width=0.5\textwidth]{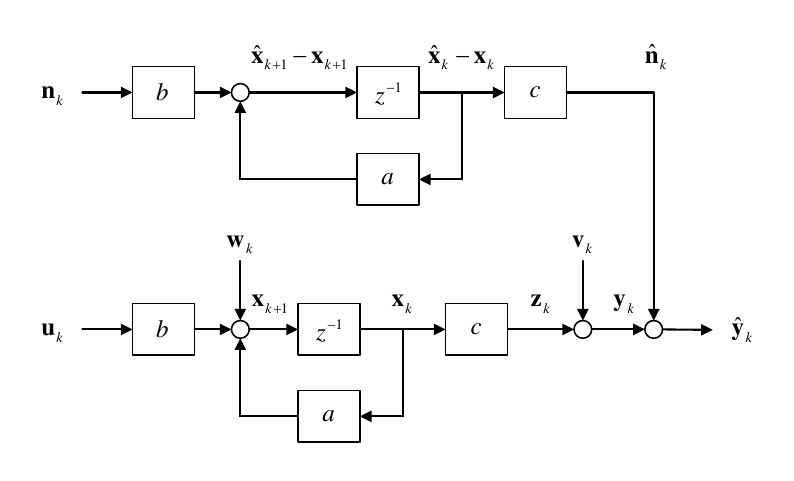}
		\vspace{-9mm}
		\caption{A Dynamical System under Injection Attack: Equivalent System.}
		\label{system3}
	\end{center}
	\vspace{-6mm}
\end{figure}

\vspace{3mm}

\begin{proof}
	To begin with, it can be verified that the power spectrum of 
	$\left\{ \mathbf{y}_{k} \right\}$  
	is given by 
	\begin{flalign}
	S_{\mathbf{y}} \left( \omega \right)
	&= \left| P \left(  \mathrm{e}^{\mathrm{j} \omega} \right) \right|^2 S_{\mathbf{u}} \left( \omega \right)
	+ \frac{1}{b^2} \left| P \left(  \mathrm{e}^{\mathrm{j} \omega} \right) \right|^2 \sigma_{\mathbf{w}}^2 + \sigma_{\mathbf{v}}^2 \nonumber \\
	&= \frac{b^2 c^2}{\left| \mathrm{e}^{\mathrm{j} \omega} - a \right|^2} S_{\mathbf{u}} \left( \omega \right)
	+ \frac{c^2}{\left| \mathrm{e}^{\mathrm{j} \omega} - a \right|^2} \sigma_{\mathbf{w}}^2 + \sigma_{\mathbf{v}}^2. \nonumber
	\end{flalign} 
	Note then that due to the  property of additivity of linear systems, the system in Fig.~\ref{system2} is equivalent to that of Fig.~\ref{system3}, where
	\begin{flalign}
	\widehat{\mathbf{y}}_{k} 
	= \mathbf{y}_{k} + \widehat{\mathbf{n}}_{k}
    , \nonumber
	\end{flalign}
	and $\left\{ \widehat{\mathbf{n}}_{k} \right\}$ is the output of the sub-system
	\begin{flalign}
	\left\{ \begin{array}{rcl}
	\widehat{\mathbf{x}}_{k+1} - \mathbf{x}_{k+1} & = & a \left( \widehat{\mathbf{x}}_{k} - \mathbf{x}_{k} \right) + b \mathbf{n}_{k},\\
	\widehat{\mathbf{n}}_{k} & = & c \left( \widehat{\mathbf{x}}_{k} - \mathbf{x}_{k} \right),
	\end{array} \right. \nonumber
	\end{flalign}
	as depicted by the upper half of Fig.~\ref{system3}; note that in this sub-system, $\left( \widehat{\mathbf{x}}_{k} - \mathbf{x}_{k} \right) \in \mathbb{R}$ is the system state, $\mathbf{n}_{k} \in \mathbb{R}$ is the system input, and $\widehat{\mathbf{n}} \in \mathbb{R}$ is the system output.
	On the other hand, the distortion constraint 
	\begin{flalign}
	\mathbb{E} \left[ \left( \widehat{\mathbf{x}}_k - \mathbf{x}_{k} \right)^{2} \right] \geq D \nonumber
	\end{flalign} 
	is then equivalent to being with a power constraint 
	\begin{flalign}
	\mathbb{E} \left[ \widehat{\mathbf{n}}_k^2 \right] \geq c^2 D,
	\nonumber
	\end{flalign} 
	since $\widehat{\mathbf{n}}_k 
	= \widehat{\mathbf{y}}_{k} - \mathbf{y}_{k}$ and thus \begin{flalign}
	\widehat{\mathbf{n}}_k^2
	= \left( \mathbf{y}_k - \widehat{\mathbf{y}}_{k} \right)^2
	= \left( c \mathbf{x}_k - c \widehat{\mathbf{x}}_{k} \right)^2
	= c^2 \left( \mathbf{x}_k - \widehat{\mathbf{x}}_{k} \right)^2.
	\nonumber
	\end{flalign}
	Accordingly, the system in Fig.~\ref{system3} may be viewed as a ``virtual channel" modeled as
	\begin{flalign}
	\widehat{\mathbf{y}}_{k} 
	= \mathbf{y}_{k} + \widehat{\mathbf{n}}_{k} \nonumber
	\end{flalign}
	with noise constraint
	\begin{flalign}
	\mathbb{E} \left[ \widehat{\mathbf{n}}_k^2 \right] \geq c^2 D,
	\nonumber
	\end{flalign} 
	where $\left\{ \mathbf{y}_k \right\}$ is the channel input, $\left\{ \widehat{\mathbf{y}}_k \right\}$ is the channel output, and $\left\{ \widehat{\mathbf{n}}_k \right\}$ is the channel noise. In addition, due to the fact that $\left\{ \mathbf{n}_{k} \right\}$ is independent of $\left\{ \mathbf{y}_{k} \right\}$, $\left\{ \widehat{\mathbf{n}}_{k} \right\}$ is also independent of $\left\{ \mathbf{y}_{k} \right\}$. 
	
	We consider first the case of a finite number of parallel (static) channels with
	\begin{flalign}
	\widehat{\mathbf{y}} = \mathbf{y} + \widehat{\mathbf{n}}, \nonumber
	\end{flalign}
	where $ \mathbf{y}, \widehat{\mathbf{y}},\widehat{\mathbf{n}} \in  \mathbb{R}^m$, and $ \widehat{\mathbf{n}} $ is independent of $ \mathbf{y} $. In addition, $\mathbf{y}$ is Gaussian with covariance $\Sigma_{\mathbf{y}}$, and the noise power constraint is given by 
	\begin{flalign} 
	\mathrm{tr} \left(\Sigma_{\widehat{\mathbf{n}} }  \right)
	= \mathbb{E} \left[ \sum_{i=1}^{m}
	\widehat{\mathbf{n}}^{2}\left( i \right) \right] \geq c^2 D, \nonumber
	\end{flalign}
	where $\widehat{\mathbf{n}} \left( i \right)$ denotes the $i$-th element of $\widehat{\mathbf{n}}$.
	In addition, according to \cite{KLproperties} (see Proposition~2 therein), we have
	\begin{flalign}
	\mathrm{KL} \left( p_{\widehat{\mathbf{y}}} \| p_{\mathbf{y}} \right)
	\geq \mathrm{KL} \left( p_{\widehat{\mathbf{y}}^{\mathrm{G}}} \| p_{\mathbf{y}} \right),
	\nonumber
	\end{flalign}
	where $\widehat{\mathbf{y}}^{\mathrm{G}}$ denotes a Gaussian random vector with the same covariance as $\widehat{\mathbf{y}}$, and equality holds if $\widehat{\mathbf{y}}$ is Gaussian. Meanwhile, it is known from Lemma~\ref{Gaussian} that
	\begin{flalign}
	\mathrm{KL} \left( p_{\widehat{\mathbf{y}}^{\mathrm{G}}} \| p_{\mathbf{y}} \right) 
	= \frac{1}{2} \left[ \mathrm{tr} \left( \Sigma_{\widehat{\mathbf{y}}} \Sigma_{\mathbf{y}}^{-1} \right) - \ln \det \left( \Sigma_{\widehat{\mathbf{y}}} \Sigma_{\mathbf{y}}^{-1} \right) - m \right]
	. \nonumber
	\end{flalign}
	On the other hand, since $\mathbf{y}$ and $\widehat{\mathbf{n}}$ are independent, we have
	\begin{flalign}
	\Sigma_{\widehat{\mathbf{y}}}
	=\Sigma_{\widehat{\mathbf{n}} +\mathbf{y} }
	=\Sigma_{\widehat{\mathbf{n}}} +\Sigma_{\mathbf{y}}. \nonumber
	\end{flalign}
	Consequently,
	\begin{flalign}
	& \mathrm{tr} \left( \Sigma_{\widehat{\mathbf{y}}} \Sigma_{\mathbf{y}}^{-1} \right) - \ln \det \left( \Sigma_{\widehat{\mathbf{y}}} \Sigma_{\mathbf{y}}^{-1} \right) \nonumber \\
	&~~~~ =  \mathrm{tr} \left[ \left( \Sigma_{\widehat{\mathbf{n}}} +\Sigma_{\mathbf{y}} \right) \Sigma_{\mathbf{y}}^{-1} \right] - \ln \det \left[ \left( \Sigma_{\widehat{\mathbf{n}}} +\Sigma_{\mathbf{y}} \right) \Sigma_{\mathbf{y}}^{-1} \right]
	. \nonumber
	\end{flalign}
	Denote the eigen-decomposition of $\Sigma_{\mathbf{y}}$ by $U_{\mathbf{y}} \Lambda_{\mathbf{y}} U^{\mathrm{T}}_{\mathbf{y}} $, where \begin{flalign}
	\Lambda_{\mathbf{y}} = \mathrm{diag} \left( \lambda_{1}, \ldots, \lambda_{m} \right). \nonumber
	\end{flalign}
	Then,
	\begin{flalign}
	&\mathrm{tr} \left[ \left( \Sigma_{\widehat{\mathbf{n}}} +\Sigma_{\mathbf{y}} \right) \Sigma_{\mathbf{y}}^{-1} \right] - \ln \det \left[ \left( \Sigma_{\widehat{\mathbf{n}}} +\Sigma_{\mathbf{y}} \right) \Sigma_{\mathbf{y}}^{-1} \right] \nonumber \\
	&~~~~ = \mathrm{tr} \left[ \left( \Sigma_{\widehat{\mathbf{n}}} + U_{\mathbf{y}} \Lambda_{\mathbf{y}} U^{\mathrm{T}}_{\mathbf{y}} \right) \left( U_{\mathbf{y}} \Lambda_{\mathbf{y}} U^{\mathrm{T}}_{\mathbf{y}} \right)^{-1} \right] \nonumber \\
    &~~~~~~~~ - \ln \det \left[ \left( \Sigma_{\widehat{\mathbf{n}}} + U_{\mathbf{y}} \Lambda_{\mathbf{y}} U^{\mathrm{T}}_{\mathbf{y}} \right) \left( U_{\mathbf{y}} \Lambda_{\mathbf{y}} U^{\mathrm{T}}_{\mathbf{y}} \right)^{-1} \right] \nonumber \\
    &~~~~ = \mathrm{tr} \left[ \left( \Sigma_{\widehat{\mathbf{n}}} + U_{\mathbf{y}} \Lambda_{\mathbf{y}} U_{\mathbf{y}}^{\mathrm{T}}  \right)  U_{\mathbf{y}} \Lambda_{\mathbf{y}}^{-1} U_{\mathbf{y}}^{\mathrm{T}}  \right] \nonumber \\
    &~~~~~~~~ - \ln \det \left[ \left( \Sigma_{\widehat{\mathbf{n}}} + U_{\mathbf{y}} \Lambda_{\mathbf{y}} U_{\mathbf{y}}^{\mathrm{T}}  \right)  U_{\mathbf{y}} \Lambda_{\mathbf{y}}^{-1} U_{\mathbf{y}}^{\mathrm{T}}  \right] \nonumber \\
    &~~~~ = \mathrm{tr} \left[ U_{\mathbf{y}} U_{\mathbf{y}}^{\mathrm{T}}  \left( \Sigma_{\widehat{\mathbf{n}}} + U_{\mathbf{y}} \Lambda_{\mathbf{y}} U_{\mathbf{y}}^{\mathrm{T}}  \right)  U_{\mathbf{y}} \Lambda_{\mathbf{y}}^{-1} U_{\mathbf{y}}^{\mathrm{T}}  \right] \nonumber \\
    &~~~~~~~~ - \ln \det \left[ U_{\mathbf{y}} U_{\mathbf{y}}^{\mathrm{T}}  \left( \Sigma_{\widehat{\mathbf{n}}} + U_{\mathbf{y}} \Lambda_{\mathbf{y}} U_{\mathbf{y}}^{\mathrm{T}}  \right)  U_{\mathbf{y}} \Lambda_{\mathbf{y}}^{-1} U_{\mathbf{y}}^{\mathrm{T}}  \right] \nonumber \\
    &~~~~ = \mathrm{tr} \left\{  U_{\mathbf{y}} 
    \left[ U_{\mathbf{y}}^{\mathrm{T}}  \left( \Sigma_{\widehat{\mathbf{n}}} + U_{\mathbf{y}} \Lambda_{\mathbf{y}} U_{\mathbf{y}}^{\mathrm{T}}  \right)  U_{\mathbf{y}} \Lambda_{\mathbf{y}}^{-1} \right] U_{\mathbf{y}}^{\mathrm{T}}   \right\} \nonumber \\
    &~~~~~~~~ - \ln \det \left\{  U_{\mathbf{y}} 
    \left[ U_{\mathbf{y}}^{\mathrm{T}}  \left( \Sigma_{\widehat{\mathbf{n}}} + U_{\mathbf{y}} \Lambda_{\mathbf{y}} U_{\mathbf{y}}^{\mathrm{T}}  \right)  U_{\mathbf{y}} \Lambda_{\mathbf{y}}^{-1} \right] U_{\mathbf{y}}^{\mathrm{T}}   \right\}\nonumber \\
    &~~~~ = \mathrm{tr}
    \left[ U_{\mathbf{y}}^{\mathrm{T}}  \left( \Sigma_{\widehat{\mathbf{n}}} + U_{\mathbf{y}} \Lambda_{\mathbf{y}} U_{\mathbf{y}}^{\mathrm{T}}  \right)  U_{\mathbf{y}} \Lambda_{\mathbf{y}}^{-1} \right] \nonumber \\
    &~~~~~~~~ - \ln \det
    \left[ U_{\mathbf{y}}^{\mathrm{T}}  \left( \Sigma_{\widehat{\mathbf{n}}} + U_{\mathbf{y}} \Lambda_{\mathbf{y}} U_{\mathbf{y}}^{\mathrm{T}}  \right)  U_{\mathbf{y}} \Lambda_{\mathbf{y}}^{-1} \right] \nonumber \\
    &~~~~ = \mathrm{tr}
    \left[  \left( U_{\mathbf{y}}^{\mathrm{T}} \Sigma_{\widehat{\mathbf{n}}} U_{\mathbf{y}}  +  \Lambda_{\mathbf{y}}  \right)   \Lambda_{\mathbf{y}}^{-1} \right] \nonumber \\
    &~~~~~~~~ - \ln \det
    \left[  \left( U_{\mathbf{y}}^{\mathrm{T}} \Sigma_{\widehat{\mathbf{n}}} U_{\mathbf{y}}  +  \Lambda_{\mathbf{y}}  \right)   \Lambda_{\mathbf{y}}^{-1} \right] \nonumber \\
	&~~~~ = \mathrm{tr} \left[ \left( \overline{\Sigma}_{\widehat{\mathbf{n}} } +  \Lambda_{\mathbf{y}}  \right) \Lambda_{\mathbf{y}}^{-1} \right] - \ln \det \left[ \left( \overline{\Sigma}_{\widehat{\mathbf{n}} } +  \Lambda_{\mathbf{y}}  \right) \Lambda_{\mathbf{y}}^{-1} \right],
	 \nonumber
	\end{flalign}
	where $\overline{\Sigma}_{\widehat{\mathbf{n}} }
	=U^{\mathrm{T}}_{\mathbf{y}}\Sigma_{\widehat{\mathbf{n}}} U_{\mathbf{y}} $.
	Denoting the diagonal terms of $\overline{\Sigma}_{\widehat{\mathbf{n}} }$ by $\overline{\sigma}_{\widehat{\mathbf{n}} \left( i \right)}^2, i=1,\ldots,m$, it is known from \cite{KLproperties} (see Proposition~4 therein) that
	\begin{flalign}
	& \mathrm{tr} \left[ \left( \overline{\Sigma}_{\widehat{\mathbf{n}} } +  \Lambda_{\mathbf{y}}  \right) \Lambda_{\mathbf{y}}^{-1} \right] - \ln \det \left[ \left( \overline{\Sigma}_{\widehat{\mathbf{n}} } +  \Lambda_{\mathbf{y}}  \right) \Lambda_{\mathbf{y}}^{-1} \right] \nonumber \\
	&~~~~ \geq \sum_{i=1}^{m} \left[ \frac{  \overline{\sigma}_{\widehat{\mathbf{n}} \left( i \right)}^2 + \lambda_{i} }{\lambda_{i}} \right]
	-  \sum_{i=1}^m \ln  \left[ \frac{ \overline{\sigma}_{\widehat{\mathbf{n}} \left( i \right)}^2 + \lambda_{i}}{\lambda_{i}} \right]
	\nonumber \\
	&~~~~ = \sum_{i=1}^{m} \left[ 1 + \frac{ \overline{\sigma}_{\widehat{\mathbf{n}} \left( i \right)}^2}{\lambda_{i}} \right]
	-  \sum_{i=1}^m \ln  \left[ 1 + \frac{ \overline{\sigma}_{\widehat{\mathbf{n}} \left( i \right)}^2 }{\lambda_{i}} \right], \nonumber
	\end{flalign}
	where equality holds if $\overline{\Sigma}_{\widehat{\mathbf{n}}}$ is diagonal. For simplicity, we denote
	\begin{flalign}
	\overline{\Sigma}_{\widehat{\mathbf{n}} }=\mathrm{diag}\left(\overline{\sigma}_{\widehat{\mathbf{n}} \left(1\right)}^2,\ldots, \overline{\sigma}_{\widehat{\mathbf{n}} \left(m\right)}^2 \right)
	=\mathrm{diag}\left(\widehat{N}_{1},\ldots,\widehat{N}_{m}\right) \nonumber
	\end{flalign}  
	when $\overline{\Sigma}_{\widehat{\mathbf{n}} }$ is diagonal.
	Then, the problem
	reduces to that of choosing $\widehat{N}_1,\ldots, \widehat{N}_m$ to minimize 
	\begin{flalign}
	\sum_{i=1}^{m} \left( 1 + \frac{ \widehat{N}_{i}}{\lambda_{i}} \right)
	-  \sum_{i=1}^m \ln  \left( 1 + \frac{ \widehat{N}_{i}}{\lambda_{i}} \right) \nonumber
	\end{flalign}
	subject to the constraint that 
	\begin{flalign}
	\sum_{i=1}^{m} \widehat{N}_{i} 
	&= \mathrm{tr} \left( \overline{\Sigma}_{\widehat{\mathbf{n}}} \right) 
	= \mathrm{tr} \left( U^{\mathrm{T}}_{\mathbf{y}}\Sigma_{\widehat{\mathbf{n}}} U_{\mathbf{y}} \right)
	= \mathrm{tr} \left( \Sigma_{\widehat{\mathbf{n}}} U_{\mathbf{y}} U^{\mathrm{T}}_{\mathbf{y}} \right) \nonumber \\
	&= \mathrm{tr} \left(  \Sigma_{\widehat{\mathbf{n}}} \right)  
	= m c^2 D. \nonumber
	\end{flalign}
	Define the Lagrange function by
	\begin{flalign}
	\sum_{i=1}^{m} \left( 1 + \frac{ \widehat{N}_{i}}{\lambda_{i}} \right)
	-  \sum_{i=1}^m \ln  \left( 1 + \frac{ \widehat{N}_{i}}{\lambda_{i}} \right) + \eta \left(\sum_{i=1}^{m} \widehat{N}_{i}- \widehat{N}\right), \nonumber
	\end{flalign}
	and differentiate it with respect to $\widehat{N}_{i}$, then we have
	\begin{flalign}
	\frac{1}{\lambda_{i}} - \frac{1}{\widehat{N}_{i}+ \lambda_{i}}
	+ \eta =0, \nonumber
	\end{flalign}
	or equivalently,
	\begin{flalign}
	\widehat{N}_{i} 
	= \frac{1}{\frac{1}{\lambda_{i}} + \eta} - \lambda_{i}
	= \frac{\lambda_{i}}{ 1 + \eta \lambda_{i}} - \lambda_{i}
	= \frac{ - \eta \lambda_{i}^2}{ 1 + \eta \lambda_{i}}, \nonumber
	\end{flalign}
	where $\eta $ satisfies 
	\begin{flalign}
	\sum_{i=1}^{m} \widehat{N}_{i} = \sum_{i=1}^{m} \frac{- \eta \lambda_{i}^2}{ 1 + \eta \lambda_{i}} = m c^2 D, \nonumber
	\end{flalign}
	while
	\begin{flalign}
    - \min_{i = 0, \ldots, m} \frac{1}{ \lambda_{i}} 
    < \eta < 0. \nonumber
	\end{flalign}
	For simplicity, we denote $\zeta = - \eta$, and accordingly,
	\begin{flalign}
	\widehat{N}_{i} 
	= \frac{ \zeta \lambda_{i}^2}{ 1 - \zeta \lambda_{i}}. \nonumber
	\end{flalign}
	where $\zeta $ satisfies 
	\begin{flalign}
	\sum_{i=1}^{m} \widehat{N}_{i} = \sum_{i=1}^{m} \frac{ \zeta \lambda_{i}^2}{ 1 - \zeta \lambda_{i}} = m c^2 D, \nonumber
	\end{flalign}
	while
	\begin{flalign}
	0 < \zeta < \min_{i = 0, \ldots, m} \frac{1}{ \lambda_{i}} . \nonumber
	\end{flalign}
	Correspondingly,
	\begin{flalign}
	&\inf_{p_{\widehat{\mathbf{n}}}} \mathrm{KL} \left( p_{\widehat{\mathbf{y}}} \| p_{\mathbf{y}} \right) \nonumber \\
	&~~~~ = \frac{1}{2} \left[ \sum_{i=1}^{m} \left( 1 + \frac{ \widehat{N}_{i}}{\lambda_{i}} \right)
	-  \sum_{i=1}^m \ln  \left( 1 + \frac{ \widehat{N}_{i}}{\lambda_{i}} \right) - m \right]
	\nonumber \\
	&~~~~ =  \sum_{i=1}^m \frac{1}{2} \left[  \frac{ \widehat{N}_{i}}{\lambda_{i}}
	-  \ln  \left( 1 + \frac{ \widehat{N}_{i}}{\lambda_{i}} \right) \right]. \nonumber
	\end{flalign}

	Consider now a scalar (dynamic) channel
	\begin{flalign}
	\widehat{\mathbf{y}}_{k} = \mathbf{y}_{k} + \widehat{\mathbf{n}}_{k}, \nonumber
	\end{flalign}
	where $ \mathbf{y}_{k}, \widehat{\mathbf{n}}_{k}, \widehat{\mathbf{y}}_{k} \in  \mathbb{R}$, while $ \left\{ \mathbf{y}_{k} \right\}$ and $ \left\{ \widehat{\mathbf{n}}_{k} \right\}$ are  independent. In addition, $\left\{ \mathbf{y}_{k} \right\}$ is stationary colored Gaussian with power spectrum $S_{\mathbf{y}} \left( \omega \right)$, whereas the noise power constraint is given by $\mathbb{E} \left[
	\widehat{\mathbf{n}}^{2}_{k} \right] \geq c^2 D$.
	We may then consider a block of consecutive uses from time $0$ to $k$ of this channel 
	as $k+1$ channels in parallel \cite{Cov:06}. Particularly, let the eigen-decomposition of $\Sigma_{\mathbf{y}_{0,\ldots,k}}$ be given by
	\begin{flalign}
	\Sigma_{\mathbf{y}_{0,\ldots,k}}=U_{\mathbf{y}_{0,\ldots,k}}\Lambda_{\mathbf{y}_{0,\ldots,k}}U^{\mathrm{T}}_{\mathbf{y}_{0,\ldots,k}}, \nonumber
	\end{flalign} 
	where
	\begin{flalign}
	\Lambda_{\mathbf{y}_{0,\ldots,k}}
	=\mathrm{diag} \left( \lambda_{0},\ldots,\lambda_{k} \right). \nonumber
	\end{flalign}
	Then, we have
	\begin{flalign}
	&\min_{p_{\widehat{\mathbf{n}}_{0,\ldots,k}}:~\sum_{i=0}^{k} \mathbb{E}
		\left[ \widehat{\mathbf{n}}_{i}^{2} \right]
		\geq \left(k+1\right)c^2 D} \frac{\mathrm{KL} \left( p_{\widehat{\mathbf{y}}_{0,\ldots,k}} \| p_{\mathbf{y}_{0,\ldots,k}} \right)}{k+1} \nonumber \\
	&~~~~ = \frac{1}{k+1} \sum_{i=0}^{k} \frac{1}{2} \left[  \frac{ \widehat{N}_{i}}{\lambda_{i}}
	-  \ln  \left( 1 + \frac{ \widehat{N}_{i}}{\lambda_{i}} \right) \right], \nonumber
	\end{flalign}
	where
	\begin{flalign}
	\widehat{N}_{i}   = \frac{\zeta \lambda_{i}^2}{ 1 - \zeta \lambda_{i}},~i=0,\ldots,k. \nonumber
	\end{flalign}
	Herein, $\zeta$ satisfies
	\begin{flalign}
	\sum_{i=0}^{k} \widehat{N}_{i}  
	= \sum_{i=0}^{k} \frac{\zeta \lambda_{i}^2}{ 1 - \zeta \lambda_{i}}
	= \left( k+1 \right) c^2 D, \nonumber
	\end{flalign}
	or equivalently,
	\begin{flalign}
	\frac{1}{k+1} \sum_{i=0}^{k} \widehat{N}_{i}  
	= \frac{1}{k+1} \left( \frac{\zeta \lambda_{i}^2}{ 1 - \zeta \lambda_{i}} \right)
	= c^2 D, \nonumber
	\end{flalign}
	while
	\begin{flalign}
	0 < \zeta < \min_{i = 0, \ldots, k} \frac{1}{ \lambda_{i}}. \nonumber
	\end{flalign}
	In addition, since the processes $ \left\{ \mathbf{y}_{k} \right\}$, $ \left\{ \widehat{\mathbf{n}}_{k} \right\}$, and $ \left\{ \widehat{\mathbf{y}}_{k} \right\}$ are stationary, we have
	\begin{flalign}
	&\lim_{k\to \infty} \min_{p_{\widehat{\mathbf{n}}_{0,\ldots,k}}:~\sum_{i=0}^{k}
		\mathbb{E}
		\left[ \widehat{\mathbf{n}}_{i}^{2} \right] \geq \left(k+1\right)c^2 D} \frac{\mathrm{KL} \left( p_{\widehat{\mathbf{y}}_{0, \ldots, k}} \| p_{\mathbf{y}_{0, \ldots, k}} \right)}{k+1} \nonumber \\
	&~~~~ = \inf_{\mathbb{E} \left[  \widehat{\mathbf{n}}_k^{2} \right] \geq c^2D} \lim_{k \to \infty} \frac{\mathrm{KL} \left( p_{\widehat{\mathbf{y}}_{0, \ldots, k}} \| p_{\mathbf{y}_{0, \ldots, k}} \right)}{k+1} \nonumber \\
	&~~~~ = \inf_{\mathbb{E} \left[ \widehat{\mathbf{n}}_k^{2} \right] \geq c^2D} \limsup_{k \to \infty} \frac{\mathrm{KL} \left( p_{\widehat{\mathbf{y}}_{0, \ldots, k}} \| p_{\mathbf{y}_{0, \ldots, k}} \right)}{k+1} \nonumber \\
	&~~~~ = \inf_{\mathbb{E} \left[ \widehat{\mathbf{n}}_k^{2} \right] \geq c^2D} \mathrm{KL}_{\infty} \left( p_{\widehat{\mathbf{y}}} \| p_{\mathbf{y}} \right) \nonumber \\
	&~~~~ = \inf_{\mathbb{E} \left[ \left( \widehat{\mathbf{x}}_k - \mathbf{x}_{k} \right)^{2} \right] \geq D} \mathrm{KL}_{\infty} \left( p_{\widehat{\mathbf{y}}} \| p_{\mathbf{y}} \right). \nonumber
	\end{flalign}
	On the other hand, since the processes are stationary, the covariance
	matrices are Toeplitz \cite{grenander1958toeplitz}, and their eigenvalues approach their limits as $k
	\to \infty$. Moreover, the densities of eigenvalues on the real line
	tend to the power spectra of the processes \cite{Pin:64, gutierrez2008asymptotically,lindquist2015linear}. Accordingly,
	\begin{flalign}
	&\inf_{\mathbb{E} \left[ \left( \widehat{\mathbf{x}}_k - \mathbf{x}_{k} \right)^{2} \right] \geq D} \mathrm{KL}_{\infty} \left( p_{\widehat{\mathbf{y}}} \| p_{\mathbf{y}} \right) \nonumber \\
	&~~~~ = \lim_{k\to \infty} \frac{1}{k+1} \sum_{i=0}^{k} \frac{1}{2} \left[  \frac{ \widehat{N}_{i}}{\lambda_{i}}
	-  \ln  \left( 1 + \frac{ \widehat{N}_{i}}{\lambda_{i}} \right) \right]
	\nonumber \\
	&~~~~ = \frac{1}{2 \pi} \int_{0}^{2 \pi} \frac{1}{2} \left\{ \frac{S_{\widehat{\mathbf{n}}} \left(\omega \right)}{S_{\mathbf{y}} \left( \omega \right)} - \ln \left[ 1 + \frac{S_{\widehat{\mathbf{n}}} \left(\omega \right)}{S_{\mathbf{y}} \left( \omega \right)} \right] \right\} \mathrm{d} \omega,
	\nonumber
	\end{flalign}
	where
	\begin{flalign}
	S_{\widehat{\mathbf{n}}} \left(\omega \right) 
	= \frac{\zeta S_{\mathbf{y}}^2 \left( \omega \right)}{ 1 - \zeta S_{\mathbf{y}} \left( \omega \right)}, \nonumber
	\end{flalign}
	and $\zeta$ satisfies
	\begin{flalign}
	\lim_{k\to \infty} \frac{1}{k+1}\sum_{i=0}^{k} \widehat{N}_{i}
	&=\frac{1}{2\pi} \int_{-\pi}^{\pi} S_{\widehat{\mathbf{n}}} \left(\omega \right) \mathrm{d}  \omega \nonumber \\ 
	&=\frac{1}{2\pi} \int_{-\pi}^{\pi} \frac{\zeta S_{\mathbf{y}}^2 \left( \omega \right)}{ 1 - \zeta S_{\mathbf{y}} \left( \omega \right)} \mathrm{d}  \omega =c^2 D, \nonumber
	\end{flalign}
	while
	\begin{flalign}
	0 < \zeta < \min_{\omega} \frac{1}{S_{\mathbf{y}} \left( \omega \right)}. \nonumber
	\end{flalign}
	
	Lastly, note that
	\begin{flalign} 
	S_{\widehat{\mathbf{n}}} \left( \omega \right)
	= \left| P \left( \mathrm{e}^{\mathrm{j} \omega} \right)  \right|^2 S_{\mathbf{n}} \left( \omega \right)
	= \frac{b^2 c^2}{\left| \mathrm{e}^{\mathrm{j} \omega} - a \right|^2} S_{\mathbf{n}} \left( \omega \right), \nonumber
	\end{flalign} 
	and hence
	\begin{flalign} 
	S_{\mathbf{n}} \left( \omega \right)
	= \frac{\left| \mathrm{e}^{\mathrm{j} \omega} - a \right|^2}{b^2 c^2}  S_{\widehat{\mathbf{n}}} \left( \omega \right)
	= \frac{\left| \mathrm{e}^{\mathrm{j} \omega} - a \right|^2}{b^2 c^2} \frac{\zeta S_{\mathbf{y}}^2 \left( \omega \right)}{ 1 - \zeta S_{\mathbf{y}} \left( \omega \right)}. \nonumber
	\end{flalign} 
	This concludes the proof.
\end{proof}


It is clear that  $S_{\mathbf{n}} \left( \omega \right)$ may be rewritten as
\begin{flalign}
S_{\mathbf{n}} \left( \omega \right)
= \frac{1}{\left| P \left( \mathrm{e}^{\mathrm{j} \omega} \right)  \right|^2} \frac{\zeta S_{\mathbf{y}}^2 \left( \omega \right)}{ 1 - \zeta S_{\mathbf{y}} \left( \omega \right)}.
\end{flalign} 
This means that the attacker only needs the knowledge of the power spectrum of the original system output $\left\{ \mathbf{y}_{k} \right\}$ and the transfer function of the system (from $\left\{ \mathbf{n}_{k} \right\}$ to $\left\{ \widehat{\mathbf{y}}_{k} \right\}$), i.e., $P \left( z \right)$, in order to carry out this worst-case attack. It is worth mentioning that the power spectrum of $\left\{ \mathbf{y}_{k} \right\}$ can be estimated based on its realizations  (see, e.g., \cite{stoica2005spectral}), while the transfer function of the system can be approximated by system identification (see, e.g., \cite{Ljung}).

On the other hand, the dual problem to that of Theorem~\ref{t1} would be: Given a certain stealthiness level in output, what is the maximum distortion in state that can be achieved by the attacker? And what is the corresponding attack? The following corollary answers these questions. 

\begin{corollary} \label{c1}
	Consider the dynamical system under injection attacks depicted in Fig.~\ref{system2}. 
	Then, in order for the attacker to ensure that the KL divergence rate between the original output and the attacked output is upper bounded by a (positive) constant $R$ as
	\begin{flalign}
	\mathrm{KL}_{\infty} \left( p_{\widehat{\mathbf{y}}} \| p_{\mathbf{y}} \right) \leq R,
	\end{flalign} 
	the maximum state distortion $\mathbb{E} \left[ \left( \widehat{\mathbf{x}}_k - \mathbf{x}_{k} \right)^{2} \right]$ that can be achieved is given by
		\begin{flalign}
	&\sup_{\mathrm{KL}_{\infty} \left( p_{\widehat{\mathbf{y}}} \| p_{\mathbf{y}} \right) \leq R} \mathbb{E} \left[ \left( \widehat{\mathbf{x}}_k - \mathbf{x}_{k} \right)^{2} \right] \nonumber \\
	&~~~~ = \frac{1}{2\pi} \int_{-\pi}^{\pi} \frac{1}{c^2} \left[  \frac{\zeta S_{\mathbf{y}}^2 \left( \omega \right)}{ 1 - \zeta S_{\mathbf{y}} \left( \omega \right)} \right] \mathrm{d}  \omega,
	\end{flalign}
	where $\zeta$ is the unique constant that satisfies
	\begin{flalign}
	\frac{1}{2 \pi} \int_{0}^{2 \pi} \frac{1}{2} \left\{ \frac{\zeta S_{\mathbf{y}} \left( \omega \right)}{1 - \zeta S_{\mathbf{y}} \left( \omega \right)} - \ln \left[  \frac{1}{1 - \zeta S_{\mathbf{y}} \left( \omega \right)} \right] \right\} \mathrm{d} \omega
	= R,
	\end{flalign} 
	while
	\begin{flalign}
	0 < \zeta < \min_{\omega} \frac{1}{S_{\mathbf{y}} \left( \omega \right)}.
	\end{flalign}
	Note that herein $S_{\mathbf{y}} \left( \omega \right)$ is given by \eqref{yspectrum}.
	Moreover, this maximum distortion is achieved when the attack signal 
	$\left\{ \mathbf{n}_{k} \right\}$ is chosen to be stationary colored Gaussian with power spectrum 	\begin{flalign}
	S_{\mathbf{n}} \left( \omega \right)
	= \frac{\left| \mathrm{e}^{\mathrm{j} \omega} - a \right|^2}{b^2 c^2} \frac{\zeta S_{\mathbf{y}}^2 \left( \omega \right)}{ 1 - \zeta S_{\mathbf{y}} \left( \omega \right)}.
	\end{flalign} 
\end{corollary}  

\vspace{3mm}

Note that the proof of Corollary~\ref{c1}, which can be carried out in a dual manner to that of Theorem~\ref{t1}, has been omitted due to lack of space.

\section{Conclusion}

In this paper, we have presented the fundamental
stealthiness-distortion tradeoffs of linear Gaussian  dynamical
systems under data injection attacks, and explicit formulas have been obtained in terms of power spectra that characterize
analytically the stealthiness-distortion tradeoffs as well as the
properties of the worst-case attacks. Potential future research directions include investigating of such tradeoffs for state estimation systems, as well as examining the security-privacy tradeoff (see, e.g., \cite{farokhi2018security, Fang21ACC2}).

\balance

\bibliographystyle{IEEEtran}
\bibliography{references}

\end{document}